\documentclass[%
twocolumn,
amsmath,amssymb,
aps,
prb,
]{revtex4-2}
\usepackage{graphicx}
\usepackage{dcolumn}
\usepackage{bm}
\usepackage{color} 
\usepackage{hyperref} 
\usepackage{xcolor}
\usepackage{multirow} 
\usepackage{booktabs}     
\usepackage{float}

\begin{document}
	
    \preprint{APS/123-QED}
    \title{
    Many-body electronic structure, self-doped double-exchange, and Hund metallicity in 1T-CrTe$_2$ bulk and monolayer
    }
    \author{Dong Hyun David Lee}
    \author{Hyeong Jun Lee}
    \author{Taek Jung Kim}
    \author{Min Yong Jeong}
    \author{Myung Joon Han}%
    \email{mj.han@kaist.ac.kr}
    \affiliation{Department of Physics, Korea Advanced Institute of Science and Technology (KAIST), Daejeon 34141, Republic of Korea}

\date{\today}

\begin{abstract}
The van der Waals (vdW) ferromagnet 1T-CrTe$_2$ is an emerging spintronics platform, notable for its high Curie temperature ($T_C$) and intriguing transport properties.
However, the fundamental interplay between the electron correlations and magnetism underlying its high $T_C$ still remains elusive. Here, using density functional theory plus dynamical mean-field theory (DFT+DMFT), we identify 1T-CrTe$_2$ as a self-doped double-exchange ferromagnet with pronounced Hund metallicity. 
This identification is grounded in the first detailed analysis of its many-body electronic structure, which reveals a dual electronic nature of Cr-$d$ orbitals where itinerant $e_g$ electrons coexist with localized $t_{2g}$ moments.
The interaction between these orbitals, mediated by Hund's coupling, drives the double-exchange ferromagnetism, establishing 1T-CrTe$_2$ as a Hund metal reminiscent of orbital-selective Mott systems. 
In the monolayer limit, while this physical picture persists, structural deformation, rather than reduced dimensionality, notably reduces $T_C$.
Our findings offer a new perspective on the high-$T_C$ ferromagnetism in 1T-CrTe$_2$, a mechanism potentially pivotal for other correlated two-dimensional vdW metallic magnets.
\end{abstract}

\maketitle

\section*{Introduction}  \label{intro} 

Two-dimensional (2D) van der Waals (vdW) magnetic materials have gained great attention for both device applications and fundamental science \cite{wang_new_2018,cortie_two-dimensional_2020,li_intrinsic_2019,kurebayashi_magnetism_2022,yang_van_2021,mak_probing_2019,seo_nearly_2020,gong_two-dimensional_2019,burch_magnetism_2018}.
Among the broader family of chromium tellurides (Cr$_{1+\delta}$Te$_2$)~\cite{saha_observation_2022, zhang_room-temperature_2023, tang_phase_2022, li_air-stable_2022, zhang_self-intercalation_2022, yang_magnetism_2023, tan_room-temperature_2023, bigi_bilayer_2025}, 1T-CrTe$_2$ is particularly compelling, featuring a high Curie temperature ($T_C$) above 300~K in its bulk form~\cite{freitas_ferromagnetism_2015} that remains robust down to the thin-film form~\cite{sun_room_2020, purbawati_-plane_2020, fabre_characterization_2021, wang_strain-_2024, zhang_substantially_2025, zhang_room-temperature_2021}. 
Furthermore, a rich variety of spintronic phenomena have been revealed in this system, spanning both experimental demonstrations and theoretical predictions~\cite{ou_zrte2crte2_2022, shi_field-free_2024, huang_colossal_2021, zhang_giant_2021, fragkos_magnetic_2022, abuawwad_mathrmcrte_2_2023, abuawwad_electrical_2024, meng_anomalous_2021, zhang_room-temperature_2021, liu_wafer-scale_2023, feng_strain-induced_2023, zhang_substantially_2025, zhang_epitaxial_2024, xu_antisymmetric_2025}.
These combined features establish 1T-CrTe$_2$ as a prime candidate for next-generation spintronic platforms.
Nevertheless, to explore its full potential, it is desirable to understand the underlying principles driving these remarkable properties.

In this study, we suggest 1T-CrTe$_2$ as a {\it self-doped} double-exchange (DE) ferromagnet, a picture grounded in a {\it dual} electronic nature of Cr-$d$ electrons.
The origin of its high-$T_c$ ferromagnetism has been a source of active debate, which has remained elusive despite previous proposals including Stoner model \cite{freitas_ferromagnetism_2015, lv_strain-controlled_2015}, superexchange \cite{lv_strain-controlled_2015, li_magnetic_2021, liu_structural_2022, wu_-plane_2022, feng_strain-induced_2023, zhu_insight_2023}, DE \cite{wang_layer_2018, wang_bethe-slater-curve-like_2020} and Ruderman-Kittel-Kasuya-Yosida (RKKY) interaction \cite{huang_evidence_2023, wang_holemediated_2024}.
By employing density functional theory plus dynamical mean-field theory (DFT+DMFT), we present the many-body electronic structure that has not been reported before. The calculated local spin susceptibility and the analysis based on `first Matsubara rule' clearly indicate the dual nature; namely, itinerant $e_g$ electrons coexist with localized $t_{2g}$ moments. Based on it, the self-doped DE picture is well established not only by the relatively weak electronegativity of Te but also by several of our synthetic simulations of varying Hund's coupling ($J_H$), coherent orbital behavior, and total energy profile.

Further, we demonstrate that 1T-CrTe$_2$ exhibits characteristic Hund metallic behaviors. 
In a class of materials recently dubbed ``Hund metals"~\cite{haule_coherenceincoherence_2009, de_medici_janus-faced_2011, yin_kinetic_2011, georges_strong_2013, georges_hund-metal_2024}, a variety of intriguing phenomena are known to be hosted, including 
unconventional superconductivity~\cite{hoshino_superconductivity_2015, lee_pairing_2018}, spin-freezing/non-Fermi-liquid behavior~\cite{werner_spin_2008, hoshino_superconductivity_2015}, spin-orbital separation (SOS)~\cite{aron_analytic_2015, stadler_dynamical_2015, deng_signatures_2019, stadler_differentiating_2021}, a broad valence histogram favoring high-spin multiplets~\cite{kutepov_self-consistent_2010, yin_kinetic_2011}, and orbital differentiation~\cite{de_medici_orbital-selective_2009, lanata_orbital_2013, fanfarillo_electronic_2015, kostin_imaging_2018, kugler_orbital_2019}. While this concept is well-established in prominent materials like iron-based superconductors~\cite{haule_coherenceincoherence_2009, yin_kinetic_2011, kostin_imaging_2018} and ruthenates~\cite{mravlje_coherence-incoherence_2011, dang_electronic_2015, kugler_strongly_2020}, its implications for 2D vdW materials remain largely unexplored~\cite{kim_fe3gete2_2022}.
Our study confirms these hallmark features in 1T-CrTe$_2$ and uncovers intriguing physics reminiscent of the orbital-selective Mott phase (OSMP), distinguishing it from canonical Hund metals.

Finally, extending our study to the monolayer limit, we find that structural deformation, rather than reduced dimensionality, is the dominant factor responsible for the notable reduction in $T_C$. 
Understanding thickness-dependent magnetic behaviors in 2D materials is crucial, as dimensionality reduction can dramatically alter magnetic properties \cite{  mak_probing_2019, burch_magnetism_2018, gong_two-dimensional_2019}. In the case of 1T-CrTe$_2$, experimental studies have revealed complex thickness-dependent phenomena: while the $T_C$ generally decreases in thinner films, an enhancement of spin polarization has also been reported \cite{zhang_room-temperature_2021, ou_zrte2crte2_2022, fabre_characterization_2021, zhang_substantially_2025}. Our analysis clarifies the interplay between structural relaxation and electronic correlation by showing how specific geometric changes in the monolayer suppress the ferromagnetic (FM) coupling, while simultaneously uncovering an enhancement of the local magnetic moment. This latter finding, originating from the persistent Hund metal nature of the system, not only explains the observed increase in spin polarization but also highlights the potential of strain engineering for controlling magnetism in correlated 2D vdW materials.

\section*{\label{res_and_disc} Results and discussions}

\subsection*{\label{mb-el-st} Many-body electronic structure by DFT+DMFT}
 
\begin{figure*}[th]
	\includegraphics[width=\textwidth]{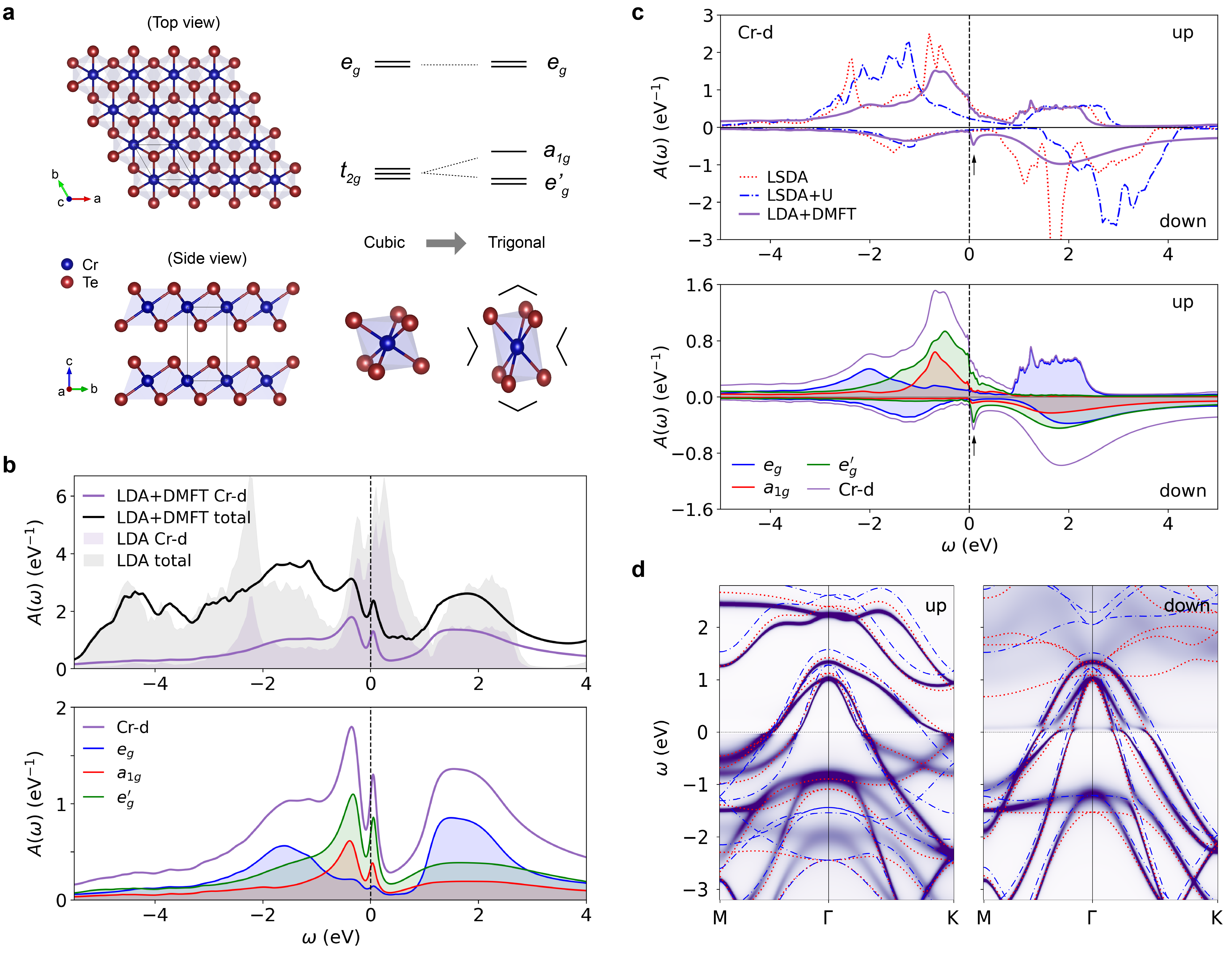}
	\caption{\label{fig1} 
    \textbf{Crystal and electronic structure of the bulk phase.}
    \textbf{a} Crystal structure of bulk 1T-CrTe$_2$ (left) and a schematic diagram of the crystal field levels (right). 
    \textbf{b} Upper panel: LDA+DMFT spectral functions $A(\omega)$ in the PM phase (solid lines; $T=450$~K) compared with the LDA results (shaded areas). 
    Lower panel: Corresponding orbital-resolved $A(\omega)$ from the LDA+DMFT calculation.
    \textbf{c} Upper panel: LDA+DMFT Cr-$d$ spectral functions in the FM phase (purple solid; $T=150$~K) compared with the LSDA (red dotted) and LSDA+$U$ (blue dash-dot) results. Positive and negative values correspond to the majority (up) and minority (down) spins, respectively.
    Lower panel: Same as in (\textbf{b}), but for the FM phase.
    \textbf{d} LDA+DMFT momentum-dependent spectral function $A(\mathbf{k}, \omega)$ at $T=150$~K for the up-spin (left panel) and down-spin (right panel) states compared with the LSDA (red dotted) and LSDA+$U$ (blue dash-dot) bands.
    }
\end{figure*}

Bulk 1T-CrTe$_2$ forms a layered hexagonal lattice with space group $P3\bar m1$ (No. 164). 
Each layer has a triangular network of edge-sharing CrTe$_6$ octahedra. These octahedra generate crystal fields that split Cr-$3d$ levels into $e_g$ and $t_{2g}$, and the latter splits further into $a_{1g}$ and $e_g'$ under trigonal distortion (Fig.~\ref{fig1}a). 
Within a simple ionic picture, 1T-CrTe$_2$ consists of Cr$^{4+}$~($3d^2$) and Te$^{2-}$~($5p^6$) ions, which yields doubly-occupied $e_g'$ and empty $a_{1g}$ and $e_g$ orbitals. In reality, however, significant $d$-$p$ hybridization makes all Cr-$d$ levels have sizable electron occupations as summarized in Table~\ref{table1}.

As a metallic magnet with partially-filled 3$d$ bands, its correlated electronic structure is of particular interest. 
While a DFT+DMFT study was recently reported, it focused on the magnetic exchange interactions, leaving the many-body electronic structure unexplored \cite{katanin_magnetic_2025}. 
Figure~\ref{fig1}b and \ref{fig1}c show our LDA+DMFT (local density approximation plus DMFT) results for paramagnetic (PM) and FM phases, respectively. 
For computational details, see Supporting Information (SI).
In the PM solution, a notable many-body feature is the quasiparticle state forming a narrow peak near the Fermi level ($E_F$; $\omega = 0$~eV) as shown in Fig.~\ref{fig1}b. Also, DMFT self-energy transfers a sizable portion of Cr-$d$ spectral weight to the higher energy region.
These characteristic features of correlated metallic systems, being clearly distinctive from those of the weakly-correlated electron approximations \cite{freitas_ferromagnetism_2015, sun_room_2020, katanin_magnetic_2025},  highlight the importance of accurate description of dynamic correlations \cite{georges_dynamical_1996, kotliar_electronic_2006}.

\begin{table}[b!] 
\caption{\label{table1}
\textbf{Calculated orbital occupations and magnetic moments.}
	Calculated orbital occupations for bulk 1T-CrTe$_2$ using L(S)DA, LSDA+$U$, and LDA+DMFT in PM (at $T= 450$~K) and FM (at $T= 150$~K) phases, along with the total magnetic moments ($M_{\text{tot}}$) for the FM phases.
    We found the results are not changed much by varying $U$ in the range of 4--7~eV with moment change $<5\%$.
 }
 \small
    \begin{tabular}{cccccccc}
    \toprule
 \multirow{2}{*}{Method} & \multirow{2}{*}{Cr-$d$} & \multirow{2}{*}{$e_g$} & \multirow{2}{*}{$t_{2g}$} & \multirow{2}{*}{($a_{1g}, e_g'$)} & \multirow{2}{*}{Te-$p$} & $M_{\text{tot}}$ \\
 & & & & & & ($\mu_B$/f.u.) \\  
    \hline                                    
    \addlinespace
     \multicolumn{7}{c}{{PM}} \\ 
     \cmidrule(lr){1-7} 
     LDA       &  4.46  & 1.70 & 2.75    &    (0.87, 1.88) & 2.12 & \multirow{2}{*}{--}\\
     DMFT      &  4.18  & 1.46 & 2.72    &    (0.91, 1.81) & 2.20  &  \\
    \hline
    \addlinespace
    \multicolumn{7}{c}{{FM}} \\ 
    \cmidrule(lr){1-7} 
     LSDA (up)  &  3.51  & 1.05 & 2.46    &    (0.83, 1.63) & 0.99 & \multirow{2}{*}{2.35} \\
        LSDA (down) &  0.86  & 0.60 & 0.26    &    (0.08, 0.18) & 1.15 \\
        LSDA+$U$ (up) &  3.66  & 0.97 & 2.68    &    (0.90, 1.79) & 1.00  & \multirow{2}{*}{2.66} \\
        LSDA+$U$ (down) &  0.56  & 0.60 & 0.15   &    (0.05,	0.10) & 1.16 \\
        DMFT (up) &  3.37  & 1.09	&  2.28   &    (0.78, 1.51) &  0.99 & \multirow{2}{*}{2.26} \\
        DMFT (down)&  0.78  & 0.37 &	0.41  & (0.13, 0.28) &  0.86 \\
        \bottomrule
        \end{tabular}
\end{table}

The result for the FM phase is shown in Fig.~\ref{fig1}c where we compare LDA+DMFT with local spin density approximation (LSDA) and LSDA$+U$ \cite{ghosh_unraveling_2023, ghosh_structural_2024}.
The calculated total magnetic moments are 2.35, 2.66, and 2.26 $\mu_B$/f.u. for LSDA, LSDA+$U$, and LDA+DMFT (at $T=150$~K), respectively, as shown in Table~\ref{table1}.
In the LSDA+$U$ solution, the static mean-field treatment of Hubbard-type correction shifts the occupied up-spin states downward and the unoccupied down-spin states upward, enhancing the spin splitting compared to  LSDA. 
In LDA+DMFT, this enhancement is moderate. Instead, as in the PM solution, the dynamical correlation effects transfer a sizable spectral weight toward the higher energies, which, for instance, manifests as a broad tail in the down-spin states above the $E_F$.

It is interesting to note that the coherent state at $\omega=0$~eV in the PM phase is absent here. While there is a small peak in the down-spin states just above $E_F$ only in LDA+DMFT (see a black arrow marked in Fig.~\ref{fig1}c), it is nearly temperature-independent. We also found that it appears only when the spin-flip Coulomb interaction term is included (see Fig.~S2), featuring the many-body effect reminiscent of the non-quasiparticle peak in half-metals \cite{katsnelson_half-metallic_2008}.

For comparison with experiments, we present the momentum-dependent spectral function $A({\bf k}, \omega)$ in Fig.~\ref{fig1}d. 
While the overall dispersion is not much different from LSDA (red dotted lines), the LDA+DMFT result exhibits substantial redistributions of spectral weights. The damping reflects the finite quasiparticle lifetime arising from electron-electron scattering. Noticeable features include the incoherent up-spin spectral weight at around $-1$~eV as well as the hole pockets centered at $\Gamma$ point and the electron-like bands near M point below $E_F$, which is in good agreement with recent angle-resolved photoemission spectroscopy (ARPES) studies \cite{zhang_room-temperature_2021, zhang_substantially_2025}.

\subsection*{Dual nature of Cr-$d$ electrons and self-doped double-exchange}
\label{Sec_double_exchange}

\begin{figure*}[th] 
\includegraphics[width=\textwidth]{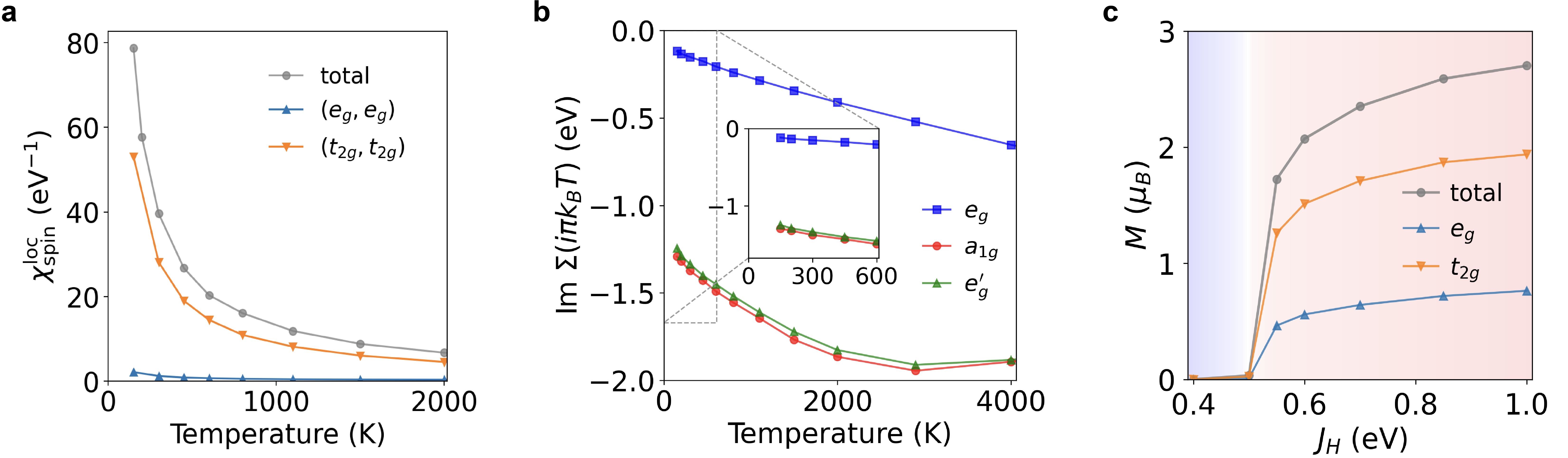}
	\caption{\label{fig2} 
    \textbf{Dual electronic nature driving double-exchange ferromagnetism.}
    \textbf{a} The temperature dependence of orbital-resolved static local spin susceptibility $\chi^{\rm loc}_{\rm spin}$ obtained in the PM phase.
    \textbf{b} The imaginary part of self-energy at the lowest fermionic Matsubara frequency ($i\omega_0 = i\pi k_B T$, where $k_B$ is Boltzmann constant) in the PM phase. The inset shows an enlarged view of the low-temperature region.
    \textbf{c} Orbital-resolved $M=\mu_B(\langle n_\uparrow\rangle -\langle n_\downarrow\rangle )$ as a function of $J_H$ at $T=150~\rm{K}$. 
    The FM phase sets in for $J_H \gtrsim 0.5$~eV.
	}
\end{figure*}

To better understand ferromagnetism in 1T-CrTe$_2$, here we suggest a new picture of {\it self-doped} DE.
We first investigate the localization versus delocalization of Cr-$3d$ electrons. Figure~\ref{fig2}a shows the calculated DMFT spin susceptibility $\chi^{\rm loc}_{\rm spin}$, indicating the Curie-Weiss (CW) behavior rather than  Pauli-like (see the grey line). In this regard, it is in contrast to Stoner ferromagnetism \cite{freitas_ferromagnetism_2015, lv_strain-controlled_2015,stoner_collective_1938}. Orbital dependent $\chi^{\rm loc}_{\rm spin}$ elucidates further details: the blue and orange line in Fig.~\ref{fig2}a is $\chi^{\rm loc}_{\rm spin}$ for $e_g$ and $t_{2g}$ orbitals, respectively. It is clearly noted that $e_g$ exhibits the notably different feature (being much closer to itinerancy) from the localized $t_{2g}$ moment \cite{haule_coherenceincoherence_2009, park_site-selective_2012, deng_signatures_2019, kim_fe3gete2_2022, kang_infinite-layer_2023}. This intriguing orbital dependency is also seen from the analysis based on ‘first-Matsubara-frequency rule’ \cite{chubukov_first-matsubara-frequency_2012}. This rule states that in Fermi-liquid (FL) regime, the imaginary part of local self-energy at the lowest Matsubara frequency is linearly proportional to temperature: $\text{Im} \Sigma(i\pi k_B T) \sim \lambda T$ where $\lambda$ is a real constant and $\Sigma$ is self-energy.
Figure~\ref{fig2}b reveals that $e_g$ orbitals nearly restore FL behavior with itinerant nature at low temperatures. 
On the other hand, $t_{2g}$ orbitals show a clear deviation: Down to the lowest temperature ($T= 150$~K) we investigated, the $T=0$ intercept remains well below 0. Thus it also supports our suggested picture of `dual' nature of Cr-$3d$ electrons in this material. In the SI, we present another analysis based on the hybridization function (see Fig.~S4).
This `dual' nature is the basis of {\it self-doped} DE.

In the conventional DE picture, the externally doped carriers mediate the localized spin moments to align ferromagnetically. For the typical case of colossal magneto-resistive manganites La$_{1-x}$D$_x$MnO$_3$ (D: divalent ion) \cite{zener_interaction_1951, anderson_considerations_1955, salamon_physics_2001}, the doped electrons occupy and hop around through $e_g$ band, and the localized $t_{2g}$ spins are ordered ferromagnetically with the aid of strong $J_H$ against the crystal field splitting.
The picture of `self-doped' DE was suggested to explain the ferromagnetism in CrO$_2$ \cite{korotin_cro_1998}. This intriguingly modified version can be valid when two different natures of Cr-$d$ states coexist. In the original paper by Korotin et al., it was argued based on the flat $d_{xy}$ band (locating below $E_F$) and the higher-lying dispersive $d_{yz \pm zx}$ with strong $d$-$p$ hybridization (going across $E_F$). Considering the increased electron affinity for the high oxidation state of Cr$^{4+}$ ion, the effective charge transfer is expected and the DE mechanism validated even without actual doping.

The aforementioned dual nature is therefore the solid ground for self-doped DE ferromagnetism in 1T-CrTe$_2$.
We note, while the original idea \cite{korotin_cro_1998} was suggested by largely relying on the band width and occupancy argument, DFT+DMFT  directly accesses the local/delocal nature from $\chi^{\rm loc}_{\rm spin}$. Given that Te has weaker electronegativity than O, 1T-CrTe$_2$ is expected to have greater $d$-$p$ hybridization and is better suited for the self-doped DE picture. 
In fact, previous ARPES studies identify Te-$p$-derived hole pockets near the $\Gamma$ point \cite{zhang_room-temperature_2021, zhang_substantially_2025, zhang_giant_2021}, which is also corroborated by X-ray absorption spectroscopy (XAS) and X-ray magnetic circular dichroism (XMCD) data reporting, respectively, the predominant Cr$^{3+}$ state and the spin moment exceeding $2 \mu_B$ \cite{zhang_room-temperature_2021, zhang_substantially_2025}.

Figure~\ref{fig2}c shows the calculated magnetization $M$ with varying $J_H$ at a fixed $T=150$~K $< T_C$. We note that non-zero $M$ is obtained only with the sufficiently large $J_H$ ($\gtrsim J_H^* \sim 0.5$ eV). Interestingly, not only $t_{2g}$ but also $e_g$ electrons carry sizable moments despite their small weight at $E_F$ and the itinerant nature discussed above. Both orbital moments share the common FM on-set value of $J_H^*$ and increase with $J_H$ while maintaining their relative ratio (Fig.~S5a). The pivotal role of $J_H$ is further corroborated by the increasing $T_C$ with $J_H$ (Fig.~S5b)~\cite{chattopadhyay_optical_2000} and the DFT+$U$ total energy calculation revealing the antiferromagnetic (AFM) to FM transition (Fig.~S6).
All these observations are well consistent with the suggested picture of self-doped DE.

\subsection*{Hund metallic behaviors}\label{Hmet} 

\begin{figure}[th]
	\includegraphics[width=0.50\textwidth]{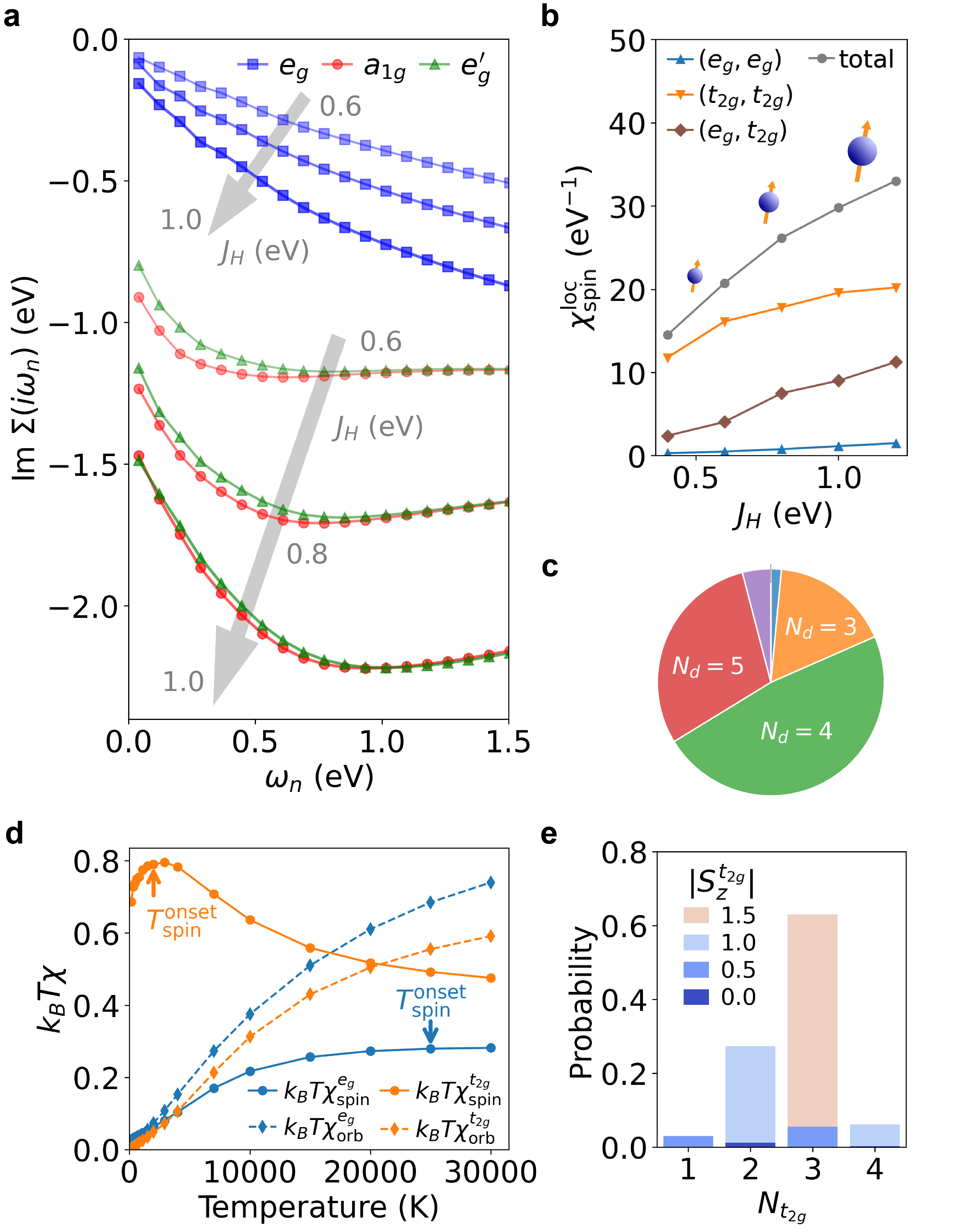}
	\caption{\label{fig3}  
\textbf{Hallmarks of Hund metallicity.}
\textbf{a} Imaginary part of self-energy on the Matsubara frequency axis $\text{Im} \Sigma(i\omega_n)$ at $T=150~\rm{K}$ and with $J_H=0.6, 0.8,$ and 1.0~eV. 
\textbf{b} The calculated local spin susceptibility $\chi^{\rm{loc}}_{\rm{spin}}$ as a function of $J_H$ at $T=450~\rm{K}$. 
\textbf{c} Probability distribution of the Cr-$d$ orbital occupancy, $N_d$. 
\textbf{d} The temperature dependence of $k_BT\chi^{\rm{loc}}_{\rm{spin/orb}}$.
The spin (solid line) and orbital (dashed line) susceptibilities $\chi$ within the $e_g$ and $t_{2g}$ manifolds are represented with blue and orange colors, respectively.
Vertical arrows mark the onset temperatures for spin screening.
\textbf{e} The valence histogram within the $t_{2g}$ manifold in terms of the electron number $N_{t_{2g}}$ and the spin-state $S^{t_{2g}}_z$.
}
\end{figure}

Let us further investigate how $J_H$ influences the electronic behavior of 1T-CrTe$_2$.
In particular, we  explore the possible `Hund metal' characters. Figure~\ref{fig3}a shows the imaginary part of self-energy $\text{Im} \Sigma(i\omega_n) $ ($\omega_n$: fermionic Matsubara frequency). 
All orbitals exhibit an enlarged $| \text{Im} \Sigma |$ with increasing $J_H$, indicative of the enhanced incoherence by $J_H$. Notably, the $t_{2g}$ self-energy is more significantly affected than the $e_g$'s. As $J_H$ approaches $0.85$~eV (presumably close to the realistic value for this system), $t_{2g}$ states enter the (so-called) `spin-frozen' regime characterized by a finite scattering rate \cite{werner_spin_2008, hoshino_superconductivity_2015}; see Fig.~S7a in SI.
The orbital-dependent incoherence can also be  quantified by scattering rate (see Fig.~S7b in SI).
These observations (i.e., the low-temperature incoherence by $J_H$ and orbital-dependent correlations) are key signatures of Hund metal \cite{georges_strong_2013}.
For the analysis of the difference between $e_g'$ and $a_{1g}$, see Fig.~S7c and S7d in SI.

The coexistence of large spin moments and significant charge fluctuations also exemplifies the Hund metal behavior \cite{haule_coherenceincoherence_2009, georges_hund-metal_2024}. 
As shown in Fig.~\ref{fig3}b, the local spin susceptibility $\chi^{\rm loc}_{\rm spin}$ increases with $J_H$, indicating the formation of large spin moments. 
This trend is further confirmed by the increased population of high-spin multiplets (Fig.~S8a). At the same time, Fig.~\ref{fig3}c shows the broad probability distribution of valence states over$N_d=3$ (16.9\%), $4$ (47.7\%) and $5$ (29.6\%), revealing significant charge fluctuations.
Its magnitude, estimated by $\langle (\Delta N_d)^2 \rangle = \langle N_d^2 \rangle - \langle N_d \rangle^2 \approx 0.67$, is comparable to that of iron-based superconductors---prototypical Hund metals \cite{moon_strong_2020}.
These fluctuations, which favor high-spin multiplets, impede the screening of local spin moments, thereby promoting Hund metallicity \cite{ryee_frozen_2023}. 
For further details, see Fig.~S8.

Next we investigate the phenomenon so-called `SOS', another hallmark of Hund metals \cite{aron_analytic_2015, stadler_dynamical_2015, deng_signatures_2019, stadler_differentiating_2021}. 
It refers to the large separation between the onset screening temperatures for spin ($T^{\rm onset}_{\rm spin}$) and orbital ($T^{\rm onset}_{\rm orb}$) degrees of freedom; i.e., $T_{\rm spin}^{\rm onset} \ll T^{\rm onset}_{\rm orb}$.
Following Ref.~\citenum{deng_signatures_2019},
we define $T^{\rm onset}_{\rm spin/orb}$ as the characteristic temperature at which $k_B T \chi$ begins to deviate from the CW behavior and to decrease ($\chi$: static local susceptibility). 
This deviation is clearly captured when we separate $\chi$ into $t_{2g}$ and $e_g$ components, as shown in Fig.~\ref{fig3}d. The SOS is well identified for $t_{2g}$ manifold. $T_{\rm spin}^{\rm onset}$ is found around 2000~K, much lower than $T^{\rm onset}_{\rm orb} > 30000$~K. 
In contrast, the screening behavior in $e_g$ manifold shows a much weaker SOS; much higher $T_{\rm spin}^{\rm onset}$ is closer to $T^{\rm onset}_{\rm orb} > 30000$~K. This substantial screening of the $e_g$ moment is consistent with the Pauli-like susceptibility discussed above.

Whereas these Hund metallic characters are well identified in 1T-CrTe$_2$, notable differences are also found. Figure~\ref{fig3}e shows the valence histogram for the $t_{2g}$ manifold. It reveals the predominance of $t_{2g}^2$ and $t_{2g}^3$, both favoring high-spin configurations ($|S^{t_{2g}}_z|=1.0 \textrm{~and~} 1.5 $, respectively) in accordance with Hund's rule. Note that the half-filling $t_{2g}^3$ configuration has higher probability than the nominally expected $t_{2g}^2$ while the latter corresponds to the typical filling for Hund metals. As a consequence, the intriguing electronic behavior is observed well distinctive from the typical Hund metals. Figure~S9a shows that  with increasing $J_{H}$, the spectral weight around $\omega \sim 0.6$~eV is gradually suppressed. The origin of this redistribution can be understood from the pronounced peak in the $t_{2g}$ self-energy, $-\mathrm{Im}\Sigma(\omega)$; see Fig.~S9b. This behavior is reminiscent of $J_H$-enhanced atomic charge gap observed in half-filled $t^3_{2g}$ systems \cite{de_medici_janus-faced_2011, georges_strong_2013}. 
Consequently, charge fluctuations are suppressed by $J_H$ (see the inset of Fig.~S9a). It contrasts with the $J_{H}$-dependence of conventional Hund metals, such as those with $t^4_{2g}$ or $t^2_{2g}$ configurations, where charge fluctuations typically increase with $J_H$ \cite{fanfarillo_electronic_2015, isidori_charge_2019}.

\subsection*{Monolayer limit} \label{Sec_monolayer}

\begin{figure}[th] 
	\includegraphics[width=0.50\textwidth]{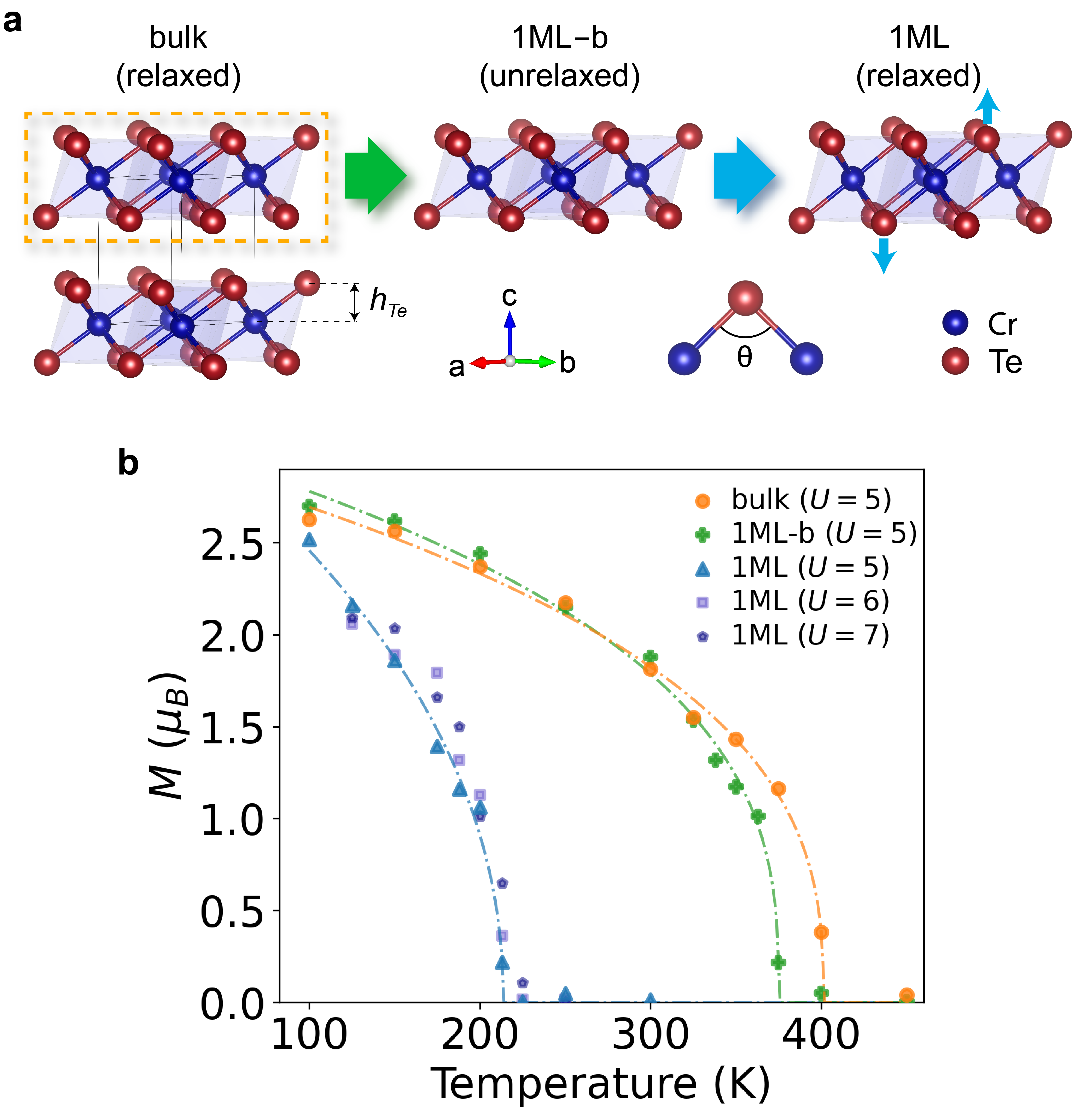}
    \caption{\label{fig4} 
    \textbf{Effect of structural deformation on monolayer magnetism.}
    \textbf{a} Structural parameters of $h_{Te}$ (the Te atom height with respect to the Cr layer) and $\theta$ (the Cr--Te--Cr bond angle). The major change by the structural relaxation is the Te atom displacements (indicated by small vertical blue arrows), which increase $h_{Te}$ and reduce $\theta$ compared to the bulk case (or equivalently 1ML-b). 
    \textbf{b} The calculated magnetization $M=\mu_B(\langle n_\uparrow\rangle -\langle n_\downarrow\rangle )$ as a function of temperature. The dash-dot lines are the fits to $U=5$~eV data with $M(T) = M_s(1-T/T_C)^\beta$. The fixed $J_H=0.85$~eV is used.
	}
\end{figure}

Finally, we investigate the monolayer limit. 
The magnetic ground state of monolayer (`1ML’) is sensitive to the in-plane lattice constant. Previous reports indicate that the smaller or distorted lattice parameter ($a \approx 3.4 \text{--} 3.7 \text{\AA}$) gives rise to AFM order \cite{xian_spin_2022, yao_ultrathin_2022, li_planar_2022, miao_tuning_2025, kushwaha_ferromagnetic_2025, armitage_electronic_2025}, whereas the larger (bulk-like; $a \approx 3.8 \text{--} 3.9 \text{\AA}$) corresponds to FM \cite{zhang_room-temperature_2021, yu_thickness-dependent_2022, wang_strain-_2024, park_unusual_2025}. As our primary focus here is on the fate of bulk ferromagnetism in the 2D limit, we adopt the bulk in-plane lattice parameter.
Figure~\ref{fig4}b shows the calculated magnetization of 1ML and bulk 1T-CrTe$_2$ (`bulk') as a function of temperature. For comparison, we also calculated the monolayer whose structure is fixed to the bulk one. Its result is indicated by `1ML-b' (green crosses). It is clearly noted that the $T_C$ is significantly reduced in 1ML ($T_C=214$~K), $\sim$53\% of the bulk value. This trend of decreasing $T_C$ by reducing the sample thickness agrees well with experiments \cite{fabre_characterization_2021, zhang_room-temperature_2021, ou_zrte2crte2_2022}; see Table~\ref{table2}. Although one previous study reports the opposite \cite{meng_anomalous_2021},
 a recent study suggested that sample-dependent variations may account for the observed difference \cite{purbawati_stability_2023}.
Considering the mean-field limitation, we estimated $T_C$ ratio between ML and bulk limit, which shows the good agreement: $T_C^{\rm mono}/T_C^{\rm bulk} = 53\%$ in our calculation and $T_C^{\rm thin}/T_C^{\rm thick}$  $\approx$ 50--67\%  in experiments \cite{zhang_room-temperature_2021, ou_zrte2crte2_2022}.

\begin{table*}[ht!]
    \caption{\label{table2}
    \textbf{Comparison of structural and magnetic properties for bulk and monolayer.}
    The calculation results of $h_{Te}$, $\theta$, $\beta$, $M_s$,  $S_{{avg}}$, and $T_C$. $\beta$, $M_s$,  and $T_C$ are obtained from the fitting of $ M(T) = M_s(1-T/T_C)^\beta $ with the results in Fig.~\ref{fig4}(b).
    The experimental $T_C$ values for bulk (thick film) and 1ML (thin film) are also listed for comparison, with the sample thickness in parentheses. 
    $S_{avg}$ (the average spin state) is calculated as the probability-weighted sum of spin states over different atomic configurations: $S_{avg} = \sum_{i} S_i  P_i$, where $i$ denotes the atomic configuration index, and $S_i$ and $P_i$ represent the corresponding spin state and probability, respectively.
     }
    \begin{ruledtabular}
        \begin{tabular}{@{}lcccccc  @{\hspace{1em}}lcccc}
        & \multicolumn{6}{c}{Calculated properties} & & \multicolumn{4}{c}{Experimental $T_C$~(K)} \\
        \cmidrule(r){2-7} \cmidrule(l){9-12} 
          & $h_{Te}$ (\AA) & $\theta$ ($^\circ$)  & $\beta$ & $M_s$ ($\mu_B$) & $S_{{avg}}$ & $T_C$ (K) & 
          & Ref. \onlinecite{ou_zrte2crte2_2022} &  Ref. \onlinecite{zhang_room-temperature_2021}  & Ref. \onlinecite{meng_anomalous_2021} & Ref. \onlinecite{fabre_characterization_2021}\\ 
            \hline  
            bulk & 1.572 & 89.4  & 0.36 & 2.99 
            \footnote{It is close to the experimentally measured one \cite{zhang_room-temperature_2021, zhang_giant_2021, zhang_substantially_2025} and theoretical results \cite{li_tunable_2021, liu_structural_2022, zhu_insight_2023, otero_fumega_controlled_2020}.} 
            & 1.65 & 401 
            \footnote{It is in reasonable agreement with experimental observations \cite{freitas_ferromagnetism_2015, sun_room_2020, purbawati_-plane_2020, zhang_room-temperature_2021, fabre_characterization_2021, ou_zrte2crte2_2022}, bearing in mind the overestimating trend of $T_C$ in mean-field approximation \cite{calderon_monte_1998, chattopadhyay_optical_2000, lichtenstein_finite-temperature_2001}.} 
            & Thick film & RT \footnote{Room temperature} (12ML) & $\sim$300 (15ML) & $\sim$187 (130 nm) & $\sim$320 (bulk) \\
            1ML-b & 1.572 & 89.4 & 0.34  & 3.09 & 1.65 & 375 
            & \multicolumn{1}{c}{--} & -- & -- & -- & -- \\
             1ML &1.658 & 87.3  & 0.47&  3.34  & 1.68 & 214 
             & Thin film  & $\gtrsim$150 (1ML) & $\sim$200 (1ML)  & $\sim$210 (10 nm)& $\sim$300 (35~nm) \\
        \end{tabular}
    \end{ruledtabular}
\end{table*}

Importantly, the calculated $T_C$ for `1ML-b' is $\sim$94\% of bulk value, demonstrating that the decreasing $T_C$ in the monolayer limit is mainly attributed to the structural effect rather than the dimensional reduction. The same trend is also found in DFT and DFT$+U$ results of Heisenberg-type exchange couplings (first neighbor $J_1$) and $T_C$ estimated from total energy  mapping and mean-field estimation, respectively.
In Table~\ref{table2}, we summarize our calculation results. It is noted that the Te height and the Cr--Te--Cr bond angle changes by +0.086 \AA~ and $-$2.1$^\circ$, respectively. 
These changes are effective enough to suppress the intralayer FM couplings by reducing the Cr-$d$ electron hopping through Te anions. 
In fact, the calculated Slater-Koster hopping parameters \cite{slater_simplified_1954} between nearest-neighbor Cr-$d$ and Te-$p$ are found to be reduced: $|V_{pd\sigma(\pi)}|=$ 1.148(0.588)~eV in 1ML-b and 1.075(0.547)~eV in 1ML.
The same feature can also be found in the density of states (DOS);
see, e.g., the bonding-antibonding splitting and the estimated bandwidth in Fig.~S10a. 
Our results suggest the strain engineering as a useful path to control $T_C$ and other magnetic properties \cite{lv_strain-controlled_2015, otero_fumega_controlled_2020,  gao_thickness_2021, li_magnetic_2021, liu_structural_2022, wu_-plane_2022, feng_strain-induced_2023, wang_strain-_2024, xu_tunable_2025}.

Hund metallic behaviors discussed above persist in the monolayer limit. The characteristic features of orbital differentiation and SOS remain similar to the bulk results, as shown in Fig.~S7e--h and Fig.~S11, respectively. 
It demonstrates Hund metallicity and dual nature are robust against the geometric changes in the 1ML structure. To investigate the possible change of Coulomb interaction in the reduced dimension \cite{zhong_electronics_2015, jang_microscopic_2019}, we repeat the calculation with the larger $U$ and find that the estimated $T_C$ does not depend much on $U$ (see Fig.~\ref{fig4}b).

One important finding in Table~\ref{table2} is that, as one approaches the monolayer limit, the moment size $M_s$ increases whereas $T_C$ decreases. To understand this intriguing behavior, we first investigate the self-energy. As presented in Fig.~S12 in SI, the magnitude of imaginary $\Sigma(i\omega_n)$ is enlarged in 1ML, indicative of the enhanced correlation. This feature is particularly pronounced for $a_{1g}$ and $e_g'$ at high frequency. 
The same feature has also been observed in the analysis of orbital- and thickness-dependent $Z(T)$, although its numerics likely requires further elaboration.
As a Hund metal, 1T-CrTe$_2$ is governed by correlations arising from $J_H$ rather than $U$. Consequently, the enhanced Hund correlation favors higher-spin configurations, giving rise to the enhanced moment $M_s$ for 1ML. The calculated multiplet histogram also shows that the higher spin configurations have more weights in the case of 1ML than 1ML-b (and bulk) as presented in Fig.~S13. It is also included in Table~\ref{table2} in terms of $S_{avg}$. 
These analyses provide a natural explanation for the unusual enhancement of spin polarization accompanied by $T_C$ reduction in the thinner 1T-CrTe$_2$. 
Our result is consistent with a recent spin-ARPES study by Zhang et al., which reports an increase in spin polarization as the film thickness is reduced \cite{zhang_substantially_2025}.

\section*{Discussion \label{disc}}

The current study suggests a new picture to understand the FM order in 1T-CrTe$_2$. 
As a DE picture, it is  distinct from the previously proposed itinerant Stoner ferromagnetism and RKKY interactions, while it can be closer to superexchange in the sense that anion-mediated hopping is a major factor \cite{freitas_ferromagnetism_2015, lv_strain-controlled_2015, lv_strain-controlled_2015, li_magnetic_2021, liu_structural_2022, wu_-plane_2022, feng_strain-induced_2023, zhu_insight_2023, wang_layer_2018, wang_bethe-slater-curve-like_2020, huang_evidence_2023, wang_holemediated_2024}. 
Although DE mechanism itself was also considered \cite{wang_layer_2018, wang_bethe-slater-curve-like_2020}, the orbital space in these studies was confined to $t_{2g}$ manifold and therefore the mixed-valency of Cr$^{3+}$($t_{2g}^3$) and Cr$^{4+}$($t_{2g}^2$) was taken into account.
It is clearly different from the self-doped DE picture in which both $t_{2g}$ and $e_g$ orbitals are essential as discussed above, and their dual nature provides the indispensable ground for FM order. In fact, FM ordering is suppressed if we define the correlated subspace to be $t_{2g}$ only (see Fig.~S14).

It is interesting to compare Hund-metallic 1T-CrTe$_2$ with orbital-selective Mott systems. While the latter has an obvious difference by having well-developed Mott gap for the specific orbital, the inter-orbital interaction also plays a crucial role in OSMP, and  $J_H$ enhances the incoherence of the metallic (wider-bandwidth) orbital state \cite{de_medici_orbital-selective_2009, jakobi_orbital-selective_2013, kim_orbital-selective_2022, biermann_non-fermi-liquid_2005}. This behavior is observed in 1T-CrTe$_2$ as shown in Fig.~S15. It is in fact responsible for the weak magnetic response within the $e_g$ sector, reflected in the subtle increase of the low-temperature spin susceptibility (see Fig.~\ref{fig2}a). The similarity can also be highlighted from the point of view of the intimacy between DE and OSMP \cite{biermann_non-fermi-liquid_2005}.

The suggested dual nature can further be corroborated by elaborate experiments. For example, a recent inelastic neutron scattering study successfully captures the dual nature of magnetic excitations and the coexistence of local and itinerant moment in a vdW metallic ferromagnet \cite{bao_neutron_2022}. Also, motivated by recent spin-ARPES studies \cite{hahn_observation_2021, zhang_substantially_2025}, we present temperature- and spin-dependent spectral function in Fig.~S16. It is noted that, upon lowering temperature, the evolution of DMFT spectral function shows some degree of orbital dependence.

\section*{\label{conc}Summary}  

In summary, we have performed the detailed investigation into the many-body electronic structure of 1T-CrTe$_2$ using DFT+DMFT, identifying it as a self-doped DE ferromagnet with pronounced Hund metallicity. This picture is built upon our discovery of a dual electronic nature in Cr-$d$ orbitals, where the analysis of local spin susceptibility and self-energy clearly distinguishes itinerant $e_g$ electrons from localized $t_{2g}$ moments. 
The overarching role of $J_H$ not only drives the DE mechanism but also establishes 1T-CrTe$_2$ as a Hund metal, with physics analogous to OSMP. 
Extending to the monolayer limit, we find that while this physical picture remains robust, structural deformation is the dominant factor suppressing the FM coupling and $T_C$, rather than the dimensional reduction itself. 
Our findings offer a unified perspective connecting DE, Hund physics, and orbital-selective correlations, providing crucial insights for designing and exploring other correlated 2D vdW metallic magnets.

\section*{Methods \label{comp}}

\subsubsection*{DFT}
We performed density functional theory (DFT) calculations using the all-electron full-potential linearized augmented plane-wave (LAPW) code WIEN2k \cite{blaha_wien2k_2020}. 
For the exchange-correlation functional, we employed the local density approximation (LDA) \cite{perdew_accurate_1992}.
We used $R_{\textrm{MT}} K_{\textrm{max}} = 7.0$ where $R_{\textrm{MT}}$ is muffin-tin radius and $K_{\textrm{max}}$ is plane-wave cutoff. The muffin-tin radii of 2.47 and 2.50 a.u. were used for Cr and Te atoms, respectively. 
We adopted the experimental lattice parameters of $a=b=3.789 ~\text{\AA}$, and $c=6.095 ~\text{\AA}$ for bulk \cite{freitas_ferromagnetism_2015}. 
The effect of vdW correction of so-called ‘Grimme D3 \cite{grimme_effect_2011}’ is found to be less than 0.02\% in terms of bond length and angle changes.
For monolayers, the in-plane lattice parameters were fixed to the bulk value, and a vacuum layer of 20 $\text{\AA}$ was taken into account. 
It is comparable with the experimental reports of Refs.~\citenum{zhang_room-temperature_2021, yu_thickness-dependent_2022, wang_strain-_2024, park_unusual_2025}.
The k-point grids of $16 \times 16 \times 8$ and $16 \times 16 \times 1$ were used for bulk and monolayers, respectively.

To analyze the orbital-dependent electron hopping, we constructed maximally localized Wannier functions (MLWFs) from the non-spin-polarized LDA band structures \cite{kunes_wien2wannier_2010, pizzi_wannier90_2020}. 
For the Wannier construction, we set an energy window from $-6.0$ to $3.0$~eV, which covers the well-isolated five Cr-$d$ and six Te-$p$ band complex from the rest. The nearest-neighbor $p$-$d$ hopping parameters  ($V_{pd\sigma}$ and $V_{pd\pi}$) were extracted by fitting Slater-Koster tight-binding matrix elements \cite{slater_simplified_1954} to the Wannier-based Hamiltonian.
To quantify the orbital-dependent bandwidths, the following  quantities were estimated with orbital index $\eta=e_g, a_{1g}, e_g'$:
\begin{equation}
\mu_\eta  =\int_{-\infty}^{\infty} dE E D_\eta(E), \\
\end{equation}
\begin{equation} \label{bw-LDA}
\sigma_\eta^2  =\int_{-\infty}^{\infty} dE \left(E-\mu_\eta\right)^2 D_\eta(E),
\end{equation}
where $D_\eta(E)$ denotes the calculated LDA density of states (DOS) at the energy $E$.
$\mu_\eta$ represents the center of mass position of DOS and $\sigma_\eta$ corresponds to the bandwidth.

\subsubsection*{DFT+$U$}
For DFT+$U$ calculations, we primarily used VASP package \cite{kresse_efficient_1996, kresse_ultrasoft_1999}.
We employed a charge-only exchange-correlation functional within Perdew-Burke-Ernzerhof (PBE) generalized gradient approximation (GGA) \cite{perdew_generalized_1996}. This choice can be crucial especially for elucidating the $J_H$ dependence \cite{ryee_effect_2018, ryee_comparative_2018, jang_charge_2018, jang_hunds_2021, park_density_2015, chen_density_2015}. The functional recipe suggested by Liechtenstein \cite{liechtenstein_density-functional_1995} was adopted with Hubbard $U=2.9$~eV and Hund $J_H=0.85$~eV. The structural parameters and the internal coordinates were optimized within this scheme. This choice of interaction parameters was the estimation from constrained RPA (cRPA) for SrCrO\textsubscript{3}, a system that shares the same formal Cr valence of $3d^2$ \cite{vaugier_hubbard_2012}. Also, it corresponds to $U_{\rm eff}=U-J_H = 2.05$~eV that is quite close to $U_{\rm eff} = 2$~eV used in previous DFT$+U$ studies for 1T-CrTe$_2$ adopting the Dudarev functional scheme \cite{sui_voltage-controllable_2017, li_tunable_2021}.

To analyze the $J_H$-dependent magnetic stability, we compared the total energies of different magnetic phases, as shown in Fig.~S6(a--d); intralayer ferromagnetic (FM), antiferromagnetic (AFM), zig-zag AFM (ZZ) order combined with interlayer FM, AFM order. 
These DFT$+U$ calculations were performed with the $2\times \sqrt{3} \times 2$ supercell and the $8 \times 10 \times 4$ k-point grid at a fixed $U$ of $2.9$~eV and the crystal structure of the bulk FM phase.
We also checked and found that the internal relaxation for each magnetic phase only slightly shifts the phase boundaries and do not change the conclusions.

The DOS in Fig.~\ref{fig1}c is obtained from the DFT+$U$ calculation using WIEN2k to maintain the consistency with the other results therein (i.e., LSDA and LDA+DMFT). Since the charge-only DFT combination with $+U$ functional is not available in this code, it is the result from LSDA+$U$ (rather than LDA$+U$). We found, however, that they are not much different from each other from the comparison to the VASP calculation result conducted within charge-only LDA+$U$.

\subsubsection*{DFT+DMFT}
To describe the dynamical electronic correlations, we employed DFT plus dynamical mean-field theory (DFT+DMFT)  \cite{georges_dynamical_1996, kotliar_electronic_2006}, which treats both itinerant and localized electrons on an equal footing.
Our charge self-consistent DFT+DMFT calculations were performed using the EDMFTF package \cite{haule_dynamical_2010}, interfaced with WIEN2k.
For the DFT part of the scheme, we used the LDA functional \cite{haule_free_2015}.
Hybridization-expansion continuous-time quantum Monte Carlo (CTQMC) was adopted for the DMFT quantum impurity solver \cite{haule_quantum_2007}. 
A large hybridization window from $-$10 to 10~eV was chosen.
To obtain one-particle spectral functions on the real-frequency axis, we performed an analytic continuation of the self-energy from the imaginary axis using the maximum entropy method \cite{haule_dynamical_2010, silver_maximum-entropy_1990, bryan_maximum_1990, jarrell_bayesian_1996, gunnarsson_analytical_2010, reymbaut_maximum_2015, bergeron_algorithms_2016,  sim_maximum_2018}.

In 1T-CrTe$_2$, each Cr atom is under an octahedral environment of six Te anions with trigonal distortion, exhibiting the $D_{3d}$ point group symmetry \cite{georgescu_trigonal_2022}. We used a trigonal basis composed of two doublets ($e_g$ and $e_g'$) and a singlet ($a_{1g}$).
As also shown in Fig.~1b (main text), $t_{2g}$ ($=a_{1g}+e_g'$) states are dominant at the Fermi level ($E_F$) while $d$-$p$ states also cross $E_F$ and finite $e_g$ states are available at $E_F$. Including $e_g$ orbitals (as well as $t_{2g}$ orbitals) as the correlated subspace in the DFT+DMFT procedure is found to be crucial as discussed in the manuscript and below.

For the local Coulomb interaction, we employed the fully rotationally invariant (`Full') form, which includes spin-flip and pair-hopping terms, contrary to another possible choice of  so-called density-density (`den-den') approximation. 
It can be crucial for the accurate description of both atomic multiplets and Kondo screening of local moments, which are central to the physics of Hund metals~\cite{georges_strong_2013, valli_kondo_2020, antipov_role_2012}. 
Retaining full rotational invariance is also known to be important for predicting realistic magnetic transition temperatures, which are often significantly overestimated by den-den approximations~\cite{antipov_role_2012, anisimov_rotationally_2012, hausoel_local_2017}. 
While the den-den interaction yields a qualitatively similar valence histogram for our system, it fails to capture crucial dynamical effects stemming from spin-flip processes, as exemplified by a spectral feature reminiscent of the non-quasiparticle peak shown in Fig.~S2 in SI.

Our analysis explores a range of interaction parameters; however, unless otherwise specified, the results presented are from $U=F^0=5$~eV and $J_H=(F^2+F^4)/14=0.85$~eV.
This choice is supported by comparisons between our theoretical results and experimental data (see 
Section ‘Many-body electronic structure by DFT+DMFT’
and Table.~\ref{table2}).
We also checked the robustness of our conclusions: variations of up to ±20\% in the interaction parameters did not significantly alter the key physical features, such as the overall FM spectral shape and the magnitude of the magnetic moment, leaving the qualitative picture unchanged.
Previous cRPA calculations \cite{jang_hunds_2021, jang_microscopic_2019} of related materials show that the $J_H$ change in thin sample limit is negligible whereas that of $U$ can be as large as $\sim$45\%. For relevant comparative investigations, see Fig.~S17 in SI. 
Furthermore, these parameters are comparable to those used for other $3d$ metallic chalcogenides \cite{kim_large_2018, yin_kinetic_2011}.
We used the double-counting scheme of  Ref.~\citenum{haule_exact_2015}.
The ferromagnetic transition temperature ($T_C$) was estimated by fitting $M(T)$ data points (Fig.~\ref{fig4}b) to the critical scaling formula $M(T) = M_s(1-T/T_C)^\beta$. With Cr $d$-occupancy  $\langle n_\sigma \rangle$ for spin $\sigma$, each $M=\mu_B(\langle n_\uparrow\rangle -\langle n_\downarrow\rangle )$ point is directly obtained from DFT+DMFT calculation.

The DMFT local susceptibilities are defined as:
\begin{equation}
\chi^{\rm loc}_{\rm spin/orb} = \int _{0}^{\beta}{\langle O_{\rm spin/orb}(\tau)O_{\rm spin/orb}(0)\rangle  d\tau}, 
\end{equation}
where $\beta = 1/(k_B T)$ is the inverse temperature ($k_B$: Boltzmann constant), and $\tau$ is an imaginary time. 
For spin susceptibility, 
$O_{\rm spin} = \frac{1}{2}\sum _{m}{[n_m^{\uparrow}(\tau)-n_m^{\downarrow}(\tau) ]}$. 
The summation in $O_{\rm spin}$ runs over orbital index $m$ within $t_{2g}$, $e_g$, and the whole $3d$ subspaces;  $\chi^{t_{2g}}_{\rm spin}$, $\chi^{e_g}_{\rm spin}$, and $\chi^{\rm total}_{\rm spin}$, respectively.
For orbital susceptibility, 
$O_{\rm orb}= \frac{1}{2}\sum _{\sigma}{[ n_{e_g'}^{\sigma}(\tau)/2 - n_{a_{1g}}^{\sigma}(\tau) ]}$
for $\chi^{t_{2g}}_{\rm orb}$ and
$O_{\rm orb}= \frac{1}{2}\sum _{\sigma}{[ n_{d_{z^2}}^{\sigma}(\tau) - n_{d_{x^2-y^2}}^{\sigma}(\tau) ]}$ for $\chi^{e_g}_{\rm orb}$, as defined in the previous studies dealing with spin-orbital separation \cite{deng_signatures_2019, ryee_hund_2021}. The summation runs over the spin degrees of freedom $\sigma = \uparrow,\downarrow $.

\section*{Data availability}
The data that support the findings of this study are available from the corresponding author upon reasonable request.





\section*{Acknowledgements}

 We thank Siheon Ryee, Fabian B. Kugler, and Philipp Werner for the fruitful discussions. 
 This study was funded by the Korea government (MSIT) through a National Research Foundation of Korea (NRF) grant (Grant Nos. RS-2025-00559042, RS-2025-02243032, and RS-2023-00253716). The funder played no role in study design, data collection, analysis and interpretation of data, or the writing of this manuscript.

\section*{Author contributions}
D.H.D.L. performed the calculations, conducted main data analysis, and
wrote the first draft. H.J.L. and T.J.K. assisted with the interpretation and validation of the results. M.Y.J. contributed to the theoretical discussions and initiation of the project. M.J.H. conceived and supervised the project. All authors discussed the results and contributed to the review and revision of the manuscript.

\section*{Competing interests}
The authors declare no competing interests.

\end{document}